\documentclass[a4paper,10pt]{article}

\usepackage{amsmath,amsfonts,amssymb,amsxtra,graphicx,epsfig}
\newtheorem{theorem}{Theorem}
\newtheorem{lemma}{Lemma}
\newtheorem{corollary}{Corollary}

\newcommand{\N}{\ensuremath{\mathbb N}}
\newcommand{\RG}{\ensuremath{\mathcal{G}(n,r)}}
\newcommand{\US}{\ensuremath{\mathcal[0,1]^2}}
\newenvironment{Proof}{\noindent{\bf 
Proof}\hspace{0.4cm}}{\par\hfill$\Box$\par}

\newcommand{\remove}[1]{}

\title{Sharp Threshold for Hamiltonicity of\\Random Geometric Graphs\thanks{Partially
supported by the EC Research Training Network
HPRN-CT-2002-00278 (COMBSTRU) and the Spanish CYCIT: TIN2004-07925-C03-01
(GRAMMARS). The first author was also supported by \emph{La distinci\'{o} per a
la promaci\'{o} de la recerca de la Generalitat de Catalunya, 2002}.}}

\author{J.D\' \i az$^1$ \qquad D.Mitsche$^2$ \qquad X.P\' erez$^1$ \smallskip \\
{\small
$^1$Llenguatges i Sistemes Inform\`{a}tics, UPC, 08034
Barcelona }\\
{\small $^2$Institut f\"ur Theoretische Informatik, ETH Z\"urich, 8092 Z\"urich} \\
{\small\tt \{diaz,xperez\}@lsi.upc.edu, dmitsche@inf.ethz.ch}}

\date{\today}

\begin{document}

\maketitle

\begin{abstract}
We show for an arbitrary $\ell_p$ norm that the property that a random geometric graph $\mathcal G(n,r)$ contains a 
Hamiltonian cycle exhibits a sharp threshold at $r=r(n)=\sqrt{\frac{\log n}{\alpha_p n}}$, where $\alpha_p$ is the 
area of the unit disk in the $\ell_p$ norm. The proof is constructive and yields a linear time algorithm for finding
a Hamiltonian cycle of $\RG$ a.a.s., provided $r=r(n)\ge\sqrt{\frac{\log n}{(\alpha_p -\epsilon)n}}$ for some fixed 
$\epsilon > 0$. 
\end{abstract}
\section{Introduction}
\label{sec:introduction}
Given a graph $G$ on $n$ vertices, a {\em Hamiltonian cycle} is a simple cycle that
visits each vertex of $G$ exactly once.
A graph is said to be {\em Hamiltonian} if it contains a Hamiltonian cycle.
The problem of given a graph, deciding if it is Hamiltonian or not is known
to be NP-complete~\cite{GareyJohnson79}. Two
known facts for the Hamiltonicity  of random graphs 
are  that almost all $d$-regular graphs ($d \geq 3$) are Hamiltonian \cite{RobWormald94}, and that in the 
$\mathcal G_{n,p}$ model if $p(n)=(\log n+\log\log n +\omega(n))/n$, then a.a.s.
$\mathcal G_{n,p}$ is Hamiltonian \cite{KomlosSzeme83} (see also chapter~8 of \cite{Bollobas01}).
Throughout this paper, ``a.a.s.'' will abbreviate {\em asymptotically
almost surely}, that is with probability tending to 1 as $n$ goes to
$\infty$.

 A {\em random geometric graph} $\RG$~\cite{Gilbert61jsiam} is a graph resulting from
placing a set  of $n$ vertices uniformly at random and independently on the unit
square $\US$, and connecting two vertices if and only if their
{\em distance} is at most the given radius $r$, the distance
depending on the type of metric being used. The two more often
used metrics are the $\ell_2$ and the $\ell_{\infty}$ norms.
In recent times, random geometric graphs have received quite a bit
of attention to model sensor networks, and  
in general ad-hoc wireless networks (see e.g.~\cite{AkyildizAl02cn}).

Random geometric graphs are the randomized version of
unit disk graphs. An undirected graph is a {\em unit disk graph}
if its vertices can be put in one-to-one correspondence with circles of 
equal radius in the plane in such a way that two vertices are joined by an edge
iff their corresponding circles intersect. W.l.o.g. it can be assumed that the radius of the circles is 1 \cite{ClackColbourn90}. The problem
of deciding if a given unit disk graph is Hamiltonian is known to be
NP-complete \cite{ItaiPapa82}.

Many properties of random geometric graphs 
have been intensively studied, both from
the theoretical and from the empirical point of view.
It is known (see~\cite{GoelRai04stoc}) that all monotone properties of $\RG$ exhibit a sharp threshold.
For the present paper, the most relevant result on random geometric
graphs is the 
connectivity threshold: in \cite{Penrose97} it is proven  that
$r=r(n)=\sqrt{\log n/(\pi n)}$ is the sharp threshold for the connectivity of $\RG$ in the $\ell_2$ norm. 
For the $\ell_{\infty}$ norm, the sharp threshold for connectivity occurs at
$r=r(n)=\sqrt{\log n/(4n)}$ (see~\cite{AppleRusso02}). 
In general, for an arbitrary $\ell_p$ norm, for some fixed $p$, $1\leq p \leq \infty$, the sharp threshold is
known to be $r=r(n)=\sqrt{\log n/(\alpha_p n)}$, where $\alpha_p$ is the area of the unit disk
in the $\ell_p$ norm (see \cite{Penrose99} and \cite{Penrose03}).
\remove {
For the present paper, the most interesting result on random geometric
graphs is the 
connectivity threshold, for $r=r(n)=\sqrt{(1-\epsilon)\ln n/(\pi
n)}$, for any constant $\epsilon>0$, in the 
$\ell_2$  norm, $\RG$ is connected aas \cite{Penrose97}. In the $\ell_{\infty}$ norm
it is known that $r=r(n)=\sqrt{(1-\epsilon)\ln n/(4n)}$, for any constant $\epsilon>0$,
$\RG$ is connected aas \cite{AppleRusso02}. For a statement in full generality see \cite{Penrose03}.
Moreover, all those thresholds are sharp
\cite{GoelRai04stoc}.

}

A natural issue to study is the existence of Hamiltonian cycles in $\RG$. Penrose in his book \cite{Penrose03}
poses it as an open problem 
whether exactly at the point where $\RG$ gets 2-connected, the
graph also becomes  Hamiltonian a.a.s. Petit in~\cite{Petit01}
proved that for $r=\omega(\sqrt{\log n/n})$, $\RG$ is Hamiltonian a.a.s. and he 
also gave a distributed algorithm to find a Hamiltonian cycle in $\RG$ with his choice of radius.
In the present paper, we find the sharp threshold for this property in any $\ell_p$ metric.
In fact, let $p$ ($1\le p\le\infty$) be arbitrary but fixed throughout the paper, and let $\mathcal G$=$\RG$ be a random geometric graph with respect to $\ell_p$.
We first show the following
\begin{theorem}
\label{main_thm}
The property that a random geometric graph $\mathcal G$=$\RG$
contains a Hamiltonian cycle exhibits a sharp threshold at $r=\sqrt{\frac{\log n}{\alpha_p n}}$, 
where $\alpha_p$ is the area of the unit disk in the $\ell_p$ norm.

More precisely, for any $\epsilon>0$,
\begin{itemize}
\item if $r= \sqrt{\frac{\log n}{(\alpha_p+\epsilon)n}}$, then a.a.s. $\mathcal G$ contains no Hamiltonian cycle,
\item if $r= \sqrt{\frac{\log n}{(\alpha_p-\epsilon)n}}$, then a.a.s. $\mathcal G$ contains a Hamiltonian cycle.
\end{itemize}
\end{theorem}
and as a corollary of the proof, we describe a linear time algorithm that finds a Hamiltonian cycle in $\RG$ a.a.s., provided
that $r\geq \sqrt{\frac{\log n}{(\alpha_p-\epsilon)n}}$ for some fixed $\epsilon > 0$. 
\section{Proof of Theorem~\ref{main_thm}}

To prove Theorem~\ref{main_thm}, note that the lower bound of the threshold is trivial. In fact, if $r= \sqrt{\frac{\log n}{(\alpha_p+\epsilon)n}}$, then 
a.a.s. $\mathcal G$ is disconnected~\cite{Penrose99} and hence it cannot contain any Hamiltonian cycle. 
To simplify the proof of the upper bound, we need some auxiliary definitions and lemmas.
In the remainder of the section, we assume that $r= \sqrt{\frac{\log n}{(\alpha_p-\epsilon)n}}$ for some fixed $\epsilon>0$, and we show that a.a.s. $\mathcal G$ contains a Hamiltonian cycle.

Let us take $y=\left\lfloor\frac{2}{r}\right\rfloor^{-1}$. Intuitively, $y$ is close to $r/2$ but slightly smaller. 
We divide $[0,1]^2$ 
into squares of side length $y$. Call this the \emph{initial tessellation} of $[0,1]^2$.
Two different squares $R$ and $S$ are defined to be \emph{friends} if they are either adjacent (i.e., they share at least one corner) or there exists at least one other square $T$ adjacent to both $R$ and $S$. Thus, each square has at most $24$ friends.
Then, we create a second and finer tessellation of $[0,1]^2$ by dividing each square into $k^2$ new squares of side 
length $y/k\sim r/(2k)$, for some large enough but fixed $k=k(\epsilon)\in\N$. 
We call this the \emph{fine tessellation} of $[0,1]^2$, and we refer to these smaller squares as \emph{cells}. 
We note that the total numbers of squares and cells are both $\Theta(1/r^2)$. Note that with probability $1$, for every
fixed $n$, any vertex will be contained in exactly one cell (and exactly one square). In the following we always assume
this.

We say that a cell is
\emph{dense}, if it contains at least $48$ vertices of $\mathcal G$. If the cell contains at least one vertex but less than $48$ vertices,
we say the cell is \emph{sparse}. If the cell contains no vertex, the cell is \emph{empty}.
Furthermore we define an \emph{animal} to be a union of cells which is topologically connected.
The \emph{size} of an animal is the number of
different cells it contains. In particular, the squares of the initial tessellation of $[0,1]^2$ are animals of size $k^2$.
An animal is called \emph{dense} if it contains at least one dense cell. If an animal contains no dense cell,
but it contains at least one vertex of $\mathcal G$, it is called \emph{sparse}. 

{From} hereinafter, all distances in $[0,1]^2$ will be taken in the $\ell_p$ metric. As usual, the distance between two sets of points $P_1$ and $P_2$ in $[0,1]^2$ is the infimum of the distances between any pair of points in $P_1$ and $P_2$.
Two cells $c_1$ and $c_2$ are said to be \emph{close} to each other if
\[
\sup_{p_1 \in c_1, p_2 \in c_2} \{\text{distance}(p_1,p_2)\} \leq r.
\]
For an arbitrary cell $c$ at distance at least $r$ from the boundary of $[0,1]^2$, let $K=K(n)$ be the number of cells which are close to $c$ and also above and to the right of $c$. Obviously, $K$ does not depend on the particular cell we chose.
\begin{figure}[ht]
\begin{center}\epsfig{file=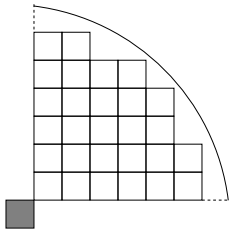, height=3cm}\end{center}
\caption{Set of cells close, above and to the right of the shaded cell}
\end{figure}
\begin{lemma}
\label{lem:k}
For any $\eta>0$, we can choose $k$ sufficiently large such that $K \geq (\alpha_p-\eta)k^2$ for $n$ large enough.
\end{lemma}
\begin{Proof}
Let $c$ be a cell at distance at least $r$ from the boundary of $[0,1]^2$. Call $A$ the union of the cells which are close to $c$ and also above and to the right of $c$. Let $p$ be the top right corner point of $c$. Define the set
\[
B = \{q\in[0,1]^2 : \text{distance}(p,q) \le r - 4y/k \}.
\]
Observe that $B\subseteq A$. Moreover, if $k$ is chosen large enough, the area of $B$ is at least $\frac14(\alpha_p-\eta) r^2$. Thus, $A$ contains at least $\frac14(\alpha_p-\eta) r^2 /(y/k)^2  \ge (\alpha_p-\eta)k^2$ cells.
\end{Proof}
\begin{lemma}\label{lem:animal}
The following statements are true a.a.s.
\begin{enumerate}
\item[i.] All animals of size $4K$ are dense.
\item[ii.] All animals of size $2K$ which touch any of the four sides of $[0,1]^2$ are dense.
\item[iii.] All cells at distance less than $4y$ from two sides of $[0,1]^2$ are dense.
\end{enumerate}
\end{lemma}
\begin{Proof}
Let $0<\delta<\epsilon$. Taking into account that the side length of each cell is $y/k\ge\frac1{2k} \sqrt\frac{\log n}{(\alpha_p-\delta)n}$ (but also $y/k \leq c\sqrt{\log n/ n}$ for some $c > 0$), the probability that any given cell is not dense (i.e. it contains at most $47$ vertices) is
\[
\sum_{i=0}^{47} \binom{n}{i} \left(\frac{y^2}{k^2}\right)^i \left(1-\frac{y^2}{k^2}\right)^{n-i}=
\Theta(1)n^{47}\left(\frac{y^2}{k^2}\right)^{47} \left(1-\frac{y^2}{k^2}\right)^{n},
\]
since the weight of this sum is concentrated in the last term. Then, plugging in the bounds for $y/k$, we get that the probability above is
\[
O(1) (n y^2/k^2)^{47} e^{-y^2n/k^2} =
O(1) (\log n)^{47} n^{-\frac{1}{4k^2(\alpha_p-\delta)}}.
\]
For each one of the cells of a given animal, we can consider the event that this particular cell is not dense. Notice that these events are negatively correlated (i.e. the probability that any particular cell is not dense conditional upon having some other cells with at most $47$ vertices is not greater than the unconditional probability). Thus, the probability that a given animal of size $4K$ contains no dense cell is at most 
\[
\left( O(1) (\log n)^{47} n^{-\frac{1}{4k^2(\alpha_p-\delta)}} \right)^{4K} = O(1) (\log n)^C n^{-\frac{K}{k^2(\alpha_p-\delta)}},
\]
for some constant $C$. Let $\rho=\frac{K}{k^2(\alpha_p-\delta)}$. From Lemma~\ref{lem:k} applied with any $0<\eta<\delta$, by choosing $k$ sufficiently large, we can guarantee that 
$\rho > 1$.
Now note that the number of animals of size $4K$ is $O(1/r^2)$ since for each
fixed shape of an animal there are $O(1/r^2)$ many choices and there is only a constant number of shapes. Thus, by 
taking a union bound over all animals and plugging in the value of $r$, we get that the probability of having an
animal without any dense cell is
\[
O(1) (\log n)^{C-1}/n^{\rho-1} = o(1),
\]
and (i) holds.

An analogous argument shows that any given animal of size $2K$ is not dense with probability
\[
O(1) (\log n)^{C/2} n^{- \rho/2}.
\]
Observe that there exist only $O(1/r)$ animals touching any of the four sides of $[0,1]^2$. Hence, the probability that one of these is not dense is
\[
O(1) (\log n)^{(C-1)/2}/ n^{(\rho-1)/2} = o(1),
\]
and (ii) is proved.

To prove (iii), we simply recall that the probability that a given cell is not dense is $o(1)$. By taking a union bound, the same argument holds for a constant number of cells.
\end{Proof}

\begin{lemma}
\label{lem:cell}
For any cell $c_1$, there exists a cell $c_2$ which is dense and close to $c_1$.
\end{lemma}
\begin{Proof}
Let $S$ be the square of the initial tessellation of $[0,1]^2$ where $c_1$ is contained, and let $A$ be the animal containing all the cells which are close to $c_1$ but different from $c_1$.
Suppose that $S$ is at distance at least $2y$ from all sides of $[0,1]^2$. Then, $A$ has size greater than $4K$, and it must contain some dense cell by Lemma~\ref{lem:animal}[i].

Otherwise, suppose that $S$ is at distance less than $2y$ from just one side of $[0,1]^2$. Then, $A$ has size greater than $2K$ and it touches one side of $[0,1]^2$, and thus it must contain some dense cell by Lemma~\ref{lem:animal}[ii].

Finally, if $S$ is at distance less than $2y$ from two sides of $[0,1]^2$, then all cells in that square must be dense by Lemma~\ref{lem:animal}[iii].
\end{Proof}
We now consider the following auxiliary graph $\mathcal G'$: the vertices of $\mathcal G'$ are all those squares belonging to the
initial tessellation of $[0,1]^2$ which are dense, and there is an edge between two dense squares $R$ and $S$ 
if they are friends and there exist cells $c_1\subset R$ and $c_2\subset S$ which are dense and close to each other. 
We observe that the maximal degree of $\mathcal G'$ is $24$. 
\remove{
We use capital letters $U$, $V$ to denote vertices of $\mathcal G'$ and reserve the lowercase $u,v,w$ for vertices of 
$\mathcal G$.
We say that a vertex $v$ of $\mathcal G$ is inside a vertex $V$ of $\mathcal G'$ if the square corresponding to $V$ contains $v$.
}
\begin{lemma}\label{lem:connected}
A.a.s., $\mathcal G'$ is connected.
\end{lemma}
\begin{Proof}
Suppose for contradiction that $\mathcal G'$ contains at least two connected components $C_1$ and $C_2$.
We denote by $D$ the union of all dense cells which are contained in some vertex (i.e., dense square) of $C_1$,
and let $H\supseteq D$ be the union of all cells which are close to some cell contained in $D$. Note that $H$ is topologically connected, and let the closed curve $\gamma$ be the outer boundary of $H$ with respect to $\mathbb R^2$.
Each connected part obtained by removing from $\gamma$ the intersection with the sides of $[0,1]^2$ is called a \emph{piece} of $\gamma$.
Define by $E$ the union of all cells in $H$ but not in $D$.
In general, $E$ might have several connected components (animals).
Moreover, all cells in $E$ must be not dense, by construction.
Note that any cell in $D$ cannot touch any piece of $\gamma$. 
Hence, each piece of $\gamma$ is touched by exactly one connected component $A\subseteq E$. Observe that, if $\gamma$ touches some side of $[0,1]^2$, then all connected components of $E$ touching some piece of $\gamma$ must also touch some side of $[0,1]^2$.

Given any of the four sides $s$ of $[0,1]^2$, the distance between $s$ and $C_1$
\remove{is defined as $\min_{S \in V(C_1)} dist(S,s)$.}
is understood to be the distance between $s$ and the dense square of $C_1$ which has the smallest distance to $s$.
We now distinguish between a few cases depending on the fact whether $C_1$ is
at distance less than $2y$ from one (or more) side(s)
of $[0,1]^2$ or not.

\noindent{\bf Case 1: } $C_1$ is at distance at least $2y$ from any side of $[0,1]^2$. \\
In this case, 
let $A$ be the only connected component of $E$ which touches $\gamma$.
Consider the uppermost dense cell $c\subset D$ (if there are several ones, choose an arbitrary one) 
and the lowermost dense cell $d\subset D$ (possibly equal to $c$). Then all cells which are close to $c$ and above $c$
and all cells which are close to $d$ and below $d$ belong to $A$. Since there are at least as many as $4K$ of these, 
we have an animal $A$ of size at least 
$4K$ without any dense cell, which by Lemma~\ref{lem:animal}[i] does not happen a.a.s.

\noindent{\bf Case 2: } $C_1$ is at distance less than $2y$ from exactly one side of $[0,1]^2$.\\
W.l.o.g. we can assume that $C_1$ is at distance less than $2y$ from the bottom side of $[0,1]^2$.
Consider the uppermost dense cell $c\subset D$ (if there are several ones, choose an arbitrary one). Let $A$ be the connected component of $E$ which contains all cells which are close to $c$ and above $c$. Note that there are at least as many as $2K$ of these cells. Moreover, $A$ touches one of the pieces of $\gamma$. Hence, we have an animal $A$ of size at least 
$2K$ without any dense cell and that touches some side of $[0,1]^2$. By Lemma~\ref{lem:animal}[ii] this does not happen a.a.s.

\noindent{\bf Case 3: } $C_1$ is at distance less than $2y$ from two opposite sides of $[0,1]^2$.\\
W.l.o.g. we can assume that $C_1$ is at distance less than $2y$ from the top and the bottom sides of $[0,1]^2$.
Among all cells contained in squares of $C_1$ that are at distance less than $4y$ from the top side of $[0,1]^2$, consider the rightmost dense cell $c$.
If $c$ is at distance less than $2y$ from that side, consider all $K$ cells which are close to $c$ and below and to the right of $c$. Otherwise, if $c$ is at distance at least $2y$ from that side, consider all $K$ cells which are close to $c$ and above and to the right of $c$.
Let $A$ be the connected component of $E$ containing these cells.
Similarly, 
among all cells contained in squares 
of $C_1$ that are at distance less than $4y$ from the bottom side of $[0,1]^2$, consider the rightmost dense cell $d$. Again, if $d$ is at distance less than $2y$ from that side, consider all $K$ cells which are close to $d$ and above and to the right of $d$. Otherwise, if $d$ is at distance at least $2y$ from that side, consider all $K$ cells which are close to $d$ and below and to the right of $d$.
Thus, in either case, we obtain $K$ cells pairwise different from the $K$ previously described ones, and let $A'$ be the connected component containing them.
$A$ and $A'$ must be the same, since they touch the same piece of $\gamma$.
Hence, we have an animal $A$ of size at least $2K$ touching at least one side of $[0,1]^2$ and without any dense cell.
By Lemma~\ref{lem:animal}[ii] this does not happen a.a.s.

\noindent{\bf Case 4: } $C_1$ is at distance less than $2y$ from one vertical and one horizontal side of $[0,1]^2$.\\
W.l.o.g. we can assume that $C_1$ is at distance less than $2y$ from the left and the top side of $[0,1]^2$.
Among all cells contained in squares of $C_1$ that are at distance less than $4y$ from the top side of $[0,1]^2$, consider the rightmost dense cell $c$.
If $c$ is at distance less than $2y$ from that side, consider all $K$ cells which are close to $c$ and below and to the right of $c$. Otherwise, if $c$ is at distance at least $2y$ from that side, consider all $K$ cells which are close to $c$ and above and to the right of $c$. Let $A$ be the connected component of $E$ containing all these $K$ cells.
By construction, all these $K$ cells are at distance less than $4y$ from the top side of $[0,1]^2$. Then,
by of Lemma~\ref{lem:animal}[iii], they must be a.a.s. at distance at least $4y$ from the left side of $[0,1]^2$, since otherwise they would be all dense.
Similarly, 
among all cells contained in squares 
of $C_1$ that are at distance less than $4y$ from the left side of $[0,1]^2$, consider the lowermost dense cell $d$. Again, if $d$ is at distance less than $2y$ from that side, consider all $K$ cells which are close to $d$ and below and to the right of $d$. Otherwise, if $d$ is at distance at least $2y$ from that side, consider all $K$ cells which are close to $d$ and below and to the left of $d$. Let $A'$ be the connected component of $E$ containing these $K$ cells.
By construction, all these $K$ cells are at distance less than $4y$ from the left side of $[0,1]^2$, and hence they must be pairwise different from the $K$ ones previously described a.a.s.
Moreover, $A$ and $A'$ must be the same, since they touch the same piece of $\gamma$.
Then we have an animal $A$ of size at least $2K$ touching at least one 
side of $[0,1]^2$ without any dense cell. By 
Lemma~\ref{lem:animal}[ii] this does not happen a.a.s.

\noindent{\bf Case 5: } $C_1$ is at distance less than $2y$ from three sides of $[0,1]^2$.\\
W.l.o.g. we can assume that $C_1$ is at distance less than $2y$ from the left, top and bottom sides of $[0,1]^2$. The argument is exactly the same
as in Case~3, and hence this case does not occur a.a.s.

In case $C_2$ is at distance at least $2y$ from some side of $[0,1]^2$, we can apply one of the above cases with $C_2$ instead of $C_1$. Thus, it suffices to consider the following:

\noindent{\bf Case 6:} Both $C_1$ and $C_2$ are at distance less than $2y$ from all four sides of $[0,1]^2$.\\
Let $Q$ be the union of all those cells at distance less than $4y$ from both the bottom and left sides of $[0,1]^2$. 
By Lemma~\ref{lem:animal}, all the cells in $Q$ must be dense, and thus must belong to squares of the same connected component of $\mathcal G'$. W.l.o.g., we can assume that they are not in $D$ (i.e. are not contained in squares of $C_1$). 
Among all cells contained in squares of $C_1$ that are at distance less than $4y$ from the bottom side of $[0,1]^2$, consider the leftmost dense cell $c$.
If $c$ is at distance less than $2y$ from that side, consider all $K$ cells which are close to $c$ and above and to the left of $c$. Otherwise, if $c$ is at distance at least $2y$ from that side, consider all $K$ cells which are close to $c$ and below and to the left of $c$. Let $A$ be the connected component of $E$ containing all these $K$ cells.
By construction, all these $K$ cells are at distance less than $4y$ from the bottom side of $[0,1]^2$. Then,
by Lemma~\ref{lem:animal}[iii], they must be a.a.s. at distance at least $4y$ from the left side of $[0,1]^2$, since otherwise they would be all dense.
Similarly, 
among all cells contained in squares 
of $C_1$ that are at distance less than $4y$ from the left side of $[0,1]^2$, consider the lowermost dense cell $d$. Again, if $d$ is at distance less than $2y$ from that side, consider all $K$ cells which are close to $d$ and below and to the right of $d$. Otherwise, if $d$ is at distance at least $2y$ from that side, consider all $K$ cells which are close to $d$ and below and to the left of $d$. Let $A'$ be the connected component of $E$ containing all these $K$ cells.
By construction, all these $K$ cells are at distance less than $4y$ from the left side of $[0,1]^2$, and hence they must be pairwise different from the $K$ ones previously described a.a.s.
Moreover, $A$ and $A'$ must be the same, since they touch the same piece of $\gamma$.
Then we have an animal $A$ of size at least $2K$ touching at least one 
side of $[0,1]^2$ without any dense cell. By 
Lemma~\ref{lem:animal}[ii] this does not happen a.a.s.
\end{Proof}

\noindent\textbf{Proof of the upper bound of Theorem~\ref{main_thm}.\hspace{0.4cm}}
Starting from $\mathcal G'$ we construct a new graph $\mathcal G''$, by adding some new vertices and edges as follows. 
Let us consider one fixed sparse square $S$ of the initial tessellation of $[0,1]^2$. 
For each sparse cell $c$ contained in $S$, we can find at least one dense cell close to it (by Lemma~\ref{lem:cell}) 
which we call the \emph{hook cell} of $c$ (if this cell is not unique, or even the square containing these cell(s) is
not unique, take an arbitrary one). This hook cell must lie inside some dense square $R$, which is a friend of $S$. 
Then, that sparse cell $c$ gets the label $R$. By grouping those ones sharing the same label, we partition the sparse 
cells of $S$ into at most $24$ groups. Each of these groups of sparse cells will be thought as a new vertex, 
added to graph $\mathcal G'$ and connected by an edge to the vertex of $\mathcal G'$ described by the common label. 
By doing this same procedure for all the remaining sparse squares, we obtain the aimed graph $\mathcal G''$. 
Those vertices in $\mathcal G''$ which already existed in $\mathcal G'$ (i.e., dense squares) are called \emph{old}, and those newly 
added ones are called \emph{new}. Notice that by construction of $\mathcal G''$ and by Lemma~\ref{lem:connected}, 
$\mathcal G''$ must be connected a.a.s.
\begin{figure}[ht]
\begin{center}\epsfig{file=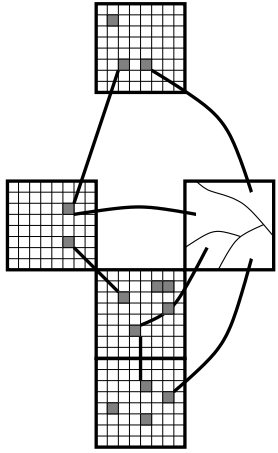, height=6cm}\end{center}
\caption{Illustration of $\mathcal G''$}
\end{figure}

Now, consider an arbitrary spanning tree $\mathcal T$ of $\mathcal G''$. Observe that the maximal degree of 
$\mathcal T$ is $24$, and that all new vertices of $\mathcal T$ have degree one and are connected to old vertices.
We use capital letters $U$, $V$ to denote vertices of $\mathcal T$ and reserve the lowercase $u,v,w$ for vertices of 
$\mathcal G$.
Fix an arbitrary traversal of $\mathcal T$ which,
starting at an arbitrary vertex, traverses each edge of $\mathcal T$ exactly twice and returns to the starting vertex. This traversal
gives an ordering in which we construct our Hamiltonian cycle in $\mathcal G$ 
(i.e., as the Hamiltonian cycle travels along the vertices of $\mathcal G$, it will visit the vertices of 
$\mathcal T$ according to this traversal).

Let us give a constructive description of our Hamiltonian cycle.
Suppose that at some time we visit an old vertex $U$ of $\mathcal T$ and that the next vertex $V$ (w.r.t. the traversal) 
is also old. Then, there must exist a pair of dense cells $c_1\subset U$, $c_2\subset V$ close to each other, and let 
$u\in c_1$ and $v\in c_2$ be vertices not used so far. In case this is not the last time we visit $U$ (w.r.t. the traversal), immediately after entering vertex $w$ inside $U$ we connect $w$ to $u$ and then $u$ is connected to $v$.
If $U$ is visited for the last time (w.r.t. the traversal), we connect from the entering vertex $w$
all vertices inside $U$ not yet used by an arbitrary Hamiltonian path (note that they form a clique in $\mathcal G$)
before leaving $U$ via $u$, and subsequently we connect $u$ to $v$.

Otherwise, suppose that at some time we visit an old vertex $U$ of $\mathcal T$ and that the next vertex $V$ 
(w.r.t. the traversal) is new.
We connect all the vertices inside $V$ (possibly just one) by an arbitrary Hamiltonian path, whose endpoints lie inside the sparse cells 
$d_1 \subset V$ and $d_2\subset V$ (possibly equal). Again this is possible since these vertices form a clique in $\mathcal G$. Let 
$c_1\subset U$ and $c_2\subset U$ (possibly equal) be the hook cells of $d_1$ and $d_2$ (i.e., $c_i$ is a dense cell in $U$ close to the sparse cell $d_i$ in $V$). Let $u\in c_1$ and $v\in c_2$ be vertices not used so far. Then, immediately after entering vertex $w$ inside $U$ we connect $w$ to $u$ and then $u$ is joined to the corresponding endpoint of the Hamiltonian path connecting the vertices inside $V$. The other endpoint is connected to $v$, and so we visit again $U$.

We observe that at some steps of the above construction we request for unused vertices of $\mathcal G$. This
is always possible: in fact, each vertex of $\mathcal T$ is visited as many times as its degree (at most $24$); for each visit of an old vertex $U$ our construction requires exactly two unused vertices $v \in c$, $w \in c$ inside some dense cell 
$c \subset U$; and $c$ contains at least $48$ vertices.
By construction, the described cycle is Hamiltonian and the result holds.

\par\hfill$\Box$\par

In the following corollary, we give an informal definition of a linear time algorithm that constructs a Hamiltonian cycle for
a specific instance of $\RG$. The procedure is based on the previous constructive proof. We assume that real arithmetic
can be done in constant time.

\begin{corollary}
\label{cor}
Let $r\ge \sqrt\frac{\log n}{(\alpha_p-\epsilon)n}$, for some fixed $\epsilon>0$. The proof of Theorem~\ref{main_thm} yields an algorithm that a.a.s. produces a Hamiltonian cycle in $\RG$ in linear time with respect to $n$.
\end{corollary}
\begin{Proof} 
Assume that the input graph satisfies all the conditions required in the proof of Theorem~\ref{main_thm}, which happens a.a.s.
Assume also that each vertex of the input graph is represented by a pair of coordinates. Observe that the total number of squares is $O(n/\log n)$, and since the number of
cells per square is constant, the same holds for the total number of cells.
%
First we compute in linear time the label of the cell and the square where each vertex is contained.
At the same time, we can find for each cell (and square) the set of vertices it contains, and mark those cells (squares) which are dense. Now, for the construction of $\mathcal G'$, note that each dense square has at most a constant number of friends to which it can be connected. Thus, the edges of $\mathcal G'$ can be obtained in time $O(n/\log n)$.
In order to construct $\mathcal G''$, for each of the $O(n/\log n)$ cells in sparse squares, we compute in constant time its hook cell and the dense square containing it.
Since both the number of vertices and the number of edges of $\mathcal G''$ are $O(n/\log n)$, we can compute in time
$O(n)$ (e.g., by Kruskal's algorithm) an arbitrary spanning tree $\mathcal T$ of $\mathcal G''$. The traversal and construction of the Hamiltonian cycle is
proportional to the number of edges in $\mathcal T$ plus the number of vertices in $\mathcal G$ and thus can be done in linear time.

\end{Proof}

\section{Conclusion and outlook}
We believe that the above construction can be generalized to obtain sharp thresholds for Hamiltonicity for random 
geometric graphs in $[0,1]^d$ ($d$ being fixed). However, it seems much more difficult to generalize the results
to arbitrary distributions of the vertices. The problem posed by Penrose~\cite{Penrose03}, whether exactly at the point
where $\RG$ gets $2$-connected the graph also becomes Hamiltonian a.a.s. or not, still remains open.

\bibliographystyle{abbrv}

\end{document}